\documentclass [11pt]{article}
\usepackage[dvips]{color}
\usepackage{epsfig}
\usepackage[normalem]{ulem}
\def\beq{\begin{equation}}
\def\eeq{\end{equation}}
\def\beqa{\begin{eqnarray}}
\def\eeqa{\end{eqnarray}}
\def\d{{\rm d}}

\begin{document}
\baselineskip0.6cm plus 1pt minus 1pt
\tolerance=1500

\begin{center}
{\LARGE\bf 
The physics of articulated toys --- a \\jumping and rotating kangaroo  }
\vskip0.4cm
{ J. G\"u\'emez$^{a,}$\footnote{guemezj@unican.es},
M. Fiolhais$^{b,}$\footnote{tmanuel@teor.fis.uc.pt}
}
\vskip0.1cm
{\it $^a$ Departamento de F\'{\i}sica Aplicada}\\ {\it Universidad de
Cantabria} \\ {\it E-39005 Santander, Spain} \\
\vskip0.1cm
{\it $^b$ Departamento de F\'\i sica and Centro de
F\'\i sica Computacional}
\\ {\it Universidade de Coimbra}
\\ {\it P-3004-516 Coimbra, Portugal}
\end{center}

\begin{abstract}
We describe the physics of an articulated toy with an internal source of energy provided by a spiral spring. The toy is a funny low cost kangaroo which jumps and rotates.  The study consists of a mechanical
and a thermodynamical analysis which makes use of the Newton and center of mass equations, the rotational equations and the first law of thermodynamics.
This amazing toy provides a nice demonstrative example how new physics insights can be  brought about when links with thermodynamics are established in the study of mechanical systems.

\end{abstract}

\section{Introduction}
\label{sec:intro}

Toys can be helpful in increasing students' motivation in the classroom. In presentations for po\-pu\-larizing and communicating science to more general audiences,
they can also help increasing the appreciation and interest in the  physical science, sometimes in such a way that everyone (especially non-scientists) will probably grasp some fundamental concepts.
However, they should be used with care: the physical description of some toys is not so easy~\cite{guemezantigos}, even in the framework of simplified models, and their usefulness
is sometimes  limited. But, at least for motivation purposes, they are always valuable~\cite{guemez09}.

In this paper we describe the motion of a toy that, due to an internal source of energy, jumps while rotates. The toy is a kangaroo but, as far as the physics description is concerned, being an object with the form of a kangaroo is just a detail, it could be something else (even a living being). Among the numerous possible objects suitable for illustration and demonstration purposes, a toy, performing on top of the instructor's table during the classroom, is definitely more likely to attract the student's attention. The accurate description of all steps of the toy's motion is intricate but some simplified assumptions are possible and meaningful. This allows us to transform the real complicated problem  into a feasible one, which is useful, in this particular case, for establishing a correspondence between the descriptions of translations and rotations, on the one hand, and, on the other hand, to bridge mechanics and thermodynamics.

The jump of the kangaroo is funny, possibly even a bit mysterious, and our aim is to apply the pertinent physical laws to describe and understand the various phases of the motion.
Though the mechanical description of rotations and translations is the result of the very same Newton's second law, students have a clear preference for translations. Since our toy performs a movement that is a combination of a translation and a rotation, it can be useful for underlining the parallelism between the mechanical treatment of each type of motion {\em per se}.  We shall assume constant forces and constant torques, therefore the real problem reduces to an almost trivial one. Nevertheless, there are some subtle points that are easy to emphasize with a simple example. In previous papers  \cite{guemez13,guemez14} we analyzed, from the mechanical and thermodynamical point of view, quite a few systems, essentially either in translation or in rotation. Here we combine both types of motion and, again, we stress the thermodynamical aspects in each phase of the motion, their similarities and asymmetries.

The  design and the manufacturing of the toy yields the kangaroo to perform a full rotation (360$^{\rm o}$) in the air while it jumps. This is because of its mass, of its shape (therefore of its moment of inertia), of the articulations between the legs and the body and also because of the power provided by the internal source of energy. The manufacturer should define and include an internal energy source suitable for the toy to perform a full turn around the centre of mass while its centre of mass raises sufficiently high  and drops down in the air. If a rotation  angle of $\sim 360^{\rm o}$ is not met, the toy doesn't work. 

Our demonstration kangaroo is a plastic three euros toy, bought in a street vendor, whose source of internal energy is a spiral spring (so, it is a low cost and very ecological item --- no batteries inside).

\begin{figure}[htb]
\begin{center}
\hspace*{-0.5cm}
\includegraphics[width=17cm]{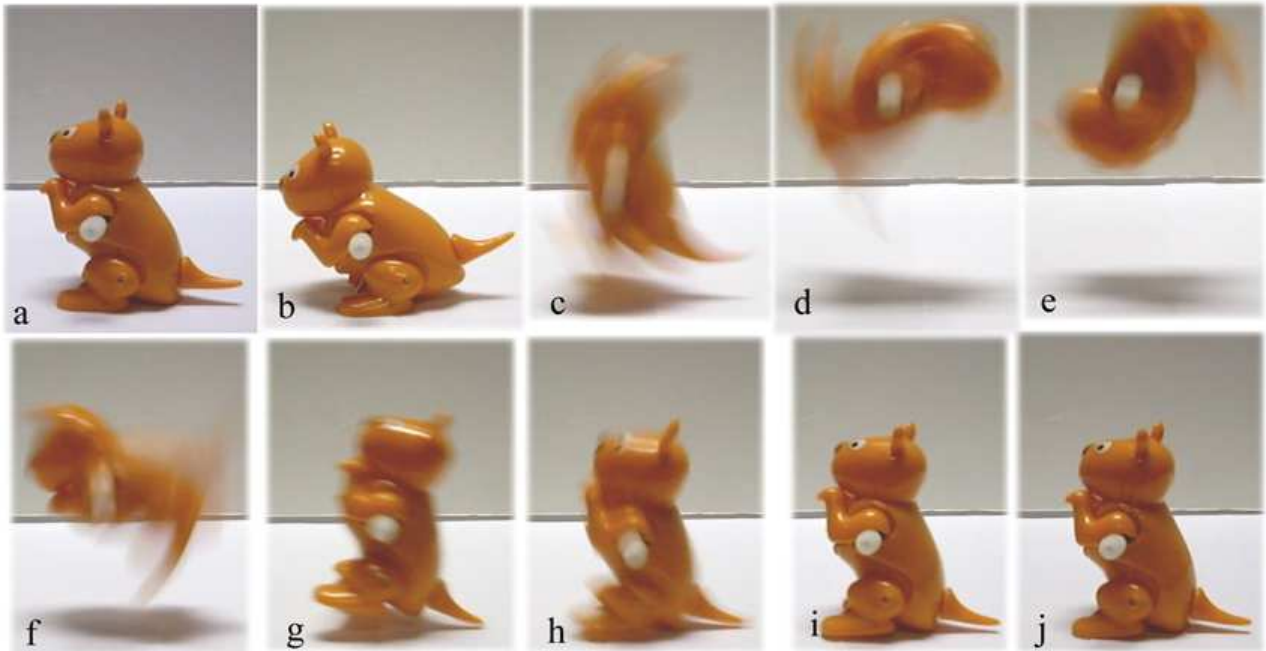}
\end{center}
\vspace*{-0.5cm}
\caption[]{\label{fig:cang1} \small The real toy performing a back somersault.  From a movie, we extracted the pictures that represent the different
phases of the motion: the preparation of the jump (a)-(b); the jump when the toy has no contact with the ground (c)-(h); and the final phase (i)-(j) when the toy stops after an initial contact
with the ground. }
\end{figure}

After providing the necessary energy to the spiral spring, we put the toy on top of a table and release it. A back somersault by the toy immediately  starts, as shown in figure \ref{fig:cang1}, and it comprises three phases:
(1) the kangaroo,  initially with the flexed legs, suddenly stretches them while rises its center of mass, increases its speed and starts rotating (a)-(b); (2) in this phase,  (c)-(h) the toy has no contact with
the ground, and rotates while its center of mass describes a parabola; (3)  this is the ``landing" phase that starts when the feet first come in contact with the ground, and it lasts until the toy completely stops  (i)-(j).

Mechanics and thermodynamics are two different branches of physics with many interrelations. However, interestingly enough,
in most university physics curricula, as well as in the high school,  thermodynamics and mechanics almost do not intercept. This is  not the case in everyday life where
both are strongly connected: a most common example is the automobile  \cite{guemez13b}, but there are many other examples  \cite{guemez13c}.
We shall see that our funny kangaroo also helps in illustrating this kind of bridging. There is no ``physics surprise" in the interpretation of the motion, we just have to apply, in combination, basic laws of mechanics and thermodynamics. With reasonable simplifying assumptions, that do not spoil the essence of the physical description, we are able to reduce the real problem to a classroom example that certainly students enjoy while they learn how basic physics principles work.

 In section 2 we briefly introduce the general formalism that will be applied in the analysis of the motion of the toy.
The discussion of the dynamics is presented in section 3 and it is essentially known. However, the subtle energetic issues related to the motion in phases 1 and 3, described in section~4, are probably less known or undervalued by instructors.  In section 5 we present the conclusions.

\section{Mechanics and thermodynamics}
Let us briefly review the basic theoretical framework needed for the mechanical and thermodynamical analyzes. A more detailed presentation can be found in \cite{guemez13,guemez14}.
For a system  of constant mass $m$, Newton's second law can be expressed by
\beq
\Delta \vec p_{\rm cm}= \int_{t_0}^t  {\vec F}_{\rm ext}\  \d t \, ,\label{eq1}
\eeq
where  ${\vec F}_{\rm ext}= \sum_i \vec f^{\rm \ ext}_i$ is the resultant external force acting upon the system and  $\Delta \vec p_{\rm cm}=m \Delta \vec v_{\rm cm}$ is the variation of the system center of mass linear momentum in the time interval $\Delta t= t-t_0$.  Equation~(\ref{eq1}) already incorporates the third law of mechanics which implies that the resultant of the internal forces vanishes,  ${\vec F}_{\rm int}=\sum_j {\vec f}^{\rm \ int}_j = \vec 0$. The above equation, expressed in a vector form, can also be equivalently given in a scalar form by means of
\beq
\Delta K_{\rm cm}= \int{\vec F}_{\rm ext}\cdot  \d {\vec r}_{\rm cm}\  \label{eq2}
\eeq
the so-called center of mass equation. On the left hand side, one has the variation of the center of mass kinetic energy ($K_{\rm cm}= {1 \over 2} \, m \, v^2_{\rm cm}$) and, on the right hand side, the
pseudo work  \cite{penchina78,sherwood83,mallin92,mungan05,jewett08v}
performed by the resultant external force. For the pseudo-work, the resultant force and the center of mass displacement should be considered [see the  integral in (\ref{eq2})], whereas for the {\em work} it is each force
and its own displacement that matters.
Equation~(\ref{eq1}), which is an integral form of Newton's equation, and the center of mass equation~(\ref{eq2}) are physically equivalent, though they use different physical magnitudes --- they express the same fundamental law of mechanics, so
they are general and do apply to all systems undergoing whatever process. When the mechanical systems perform rotations, other forms of Newton's fundamental law are better suited
such as \cite{varios,tipler04}
\beq
\Delta L=I\Delta \omega = \int_{t_0}^t  \tau_{\rm ext}\  \d t \ \ \ \ \ \ \ \  \ \Delta K_{\rm rot}= {1 \over 2} I \Delta \omega^2 = \int \tau_{\rm ext} \d \phi  \label{eq3}
\eeq
for the rotation of a system of constant moment of inertia $I$ around a principal axis of inertia containing the center of mass. The system is acted upon by an external torque, of magnitude $\tau_{\rm ext}$, whose direction is along the rotation axis. These equations are simplified versions of the most general ones, and in~(\ref{eq3}) one does not need to consider the vector character of the angular momentum, $\vec L$, of the angular velocity, $\vec \omega$, or of the torque, $\vec \tau_{\rm ext}$.  The two equations (\ref{eq3}) for the rotation, together with (\ref{eq1}) and (\ref{eq2}), for the translation, are the pertinent ones for the mechanical description of the motion presented in the next section.

A different physical law,  also applicable to any system and to any process, is the first law of thermodynamics which, incidentally,  also involves typical mechanical quantities.
That principle is a statement on energy conservation and it can be expressed by the equation
\beq
\Delta K_{\rm cm}+ \Delta U= W_{\rm ext} + Q \, . \label{totale}
\eeq
On the left hand side, in addition to the variation of the center of mass kinetic energy, one has the variation of the internal energy and, on the right hand side,
the energy fluxes to/from the system are expressed. In other words, the left hand side of equation (\ref{totale}) refers to the total energy variation of the system, whereas the right hand side expresses the energy
that crosses the system's boundary  i.e. the energy that enters or leaves the system through its boundary. This energy transfer is the sum of two contributions: the {\em external work} --- i.e. the sum of the works
performed by each external force, $w^{\rm ext}_i=\int \vec f^{\rm \ ext}_i\cdot \d \vec r_i$ --- which is given by  $W_{\rm ext}=\sum_i w^{\rm ext}_i$; and $Q$, that is the heat flow to/from the surroundings. If either $W_{\rm ext}$ or $Q$ are positive, that means an energy transfer to the system, leading to an increase of the left hand side of (\ref{totale}); if any of them is negative, that means an energy flow to the surroundings with a corresponding decrease of the total energy of the system.  It is worth noticing that, whereas the pseudo-work of the resultant external force leads to the variation of the centre of mass kinetic energy, as stated by equation (\ref{eq2}), the real work, together with the heat, may change both that kinetic energy and/or the internal energy of the system, as stated by equation (\ref{totale}).
By expressing the first law of thermodynamics in terms of equation (\ref{totale}) one implicitly assumes that $\Delta U$ includes {\em all}  energy variations that may contribute to the total internal energy variation of the system.
Those include the variations of rotational [such as $\Delta K_{\rm rot}$ as given by equation~(\ref{eq3})] and translational kinetic energies with respect to the center of mass, in addition to the variations due to temperature changes, variations of internal chemical energy (associated with chemical reactions), work performed by internal forces, etc. Of course, any process should also coply with the second law, besides the first law of thermodynamics.

We stress that equations (\ref{eq2}) and  (\ref{totale}) provide, in general, complementary information, since they correspond to two distinct fundamental laws of nature.
The study of the movement of the toy described in the next sections illustrates that complementarity.

\section{Back somersault by a ``kangaroo"}

\label{sec:cangaroo}

\begin{figure}[hbt]
\begin{center}
\hspace*{-0.5cm}
\includegraphics[width=16cm]{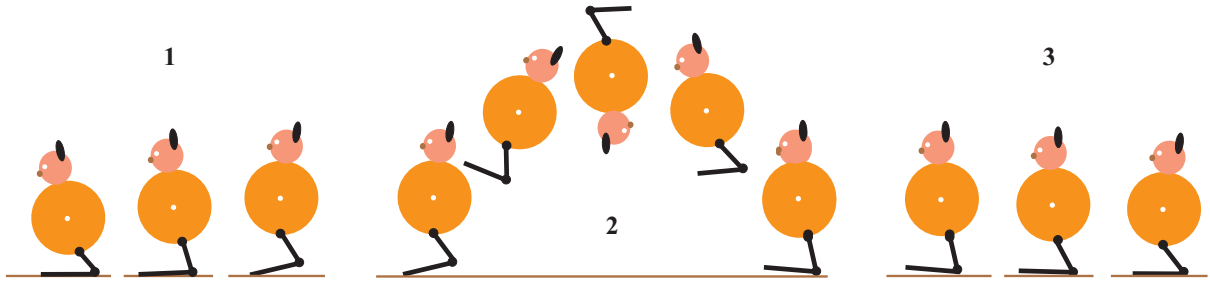}
\end{center}
\vspace*{-0.5cm}
\caption[]{\label{fig:cang2} \small Diagrammatic description of the three phases of the motion.
 In the first phase (1) there is contact with the ground while the centre of mass moves horizontally and vertically; in phase (2) the centre of mass describes a parabola while the toy performs a 360$^{\rm o}$ turn around; the final phase (3) is the deceleration phase.}
\end{figure}

We have shown, in figure \ref{fig:cang1}, the real toy performing the back somersault that comprises three phases.
In figure \ref{fig:cang2} we illustrate pictorially these three phases of the kangaroo's motion.

\begin{figure}[htb]
\begin{center}
\hspace*{0.5cm}
\includegraphics[width=11cm]{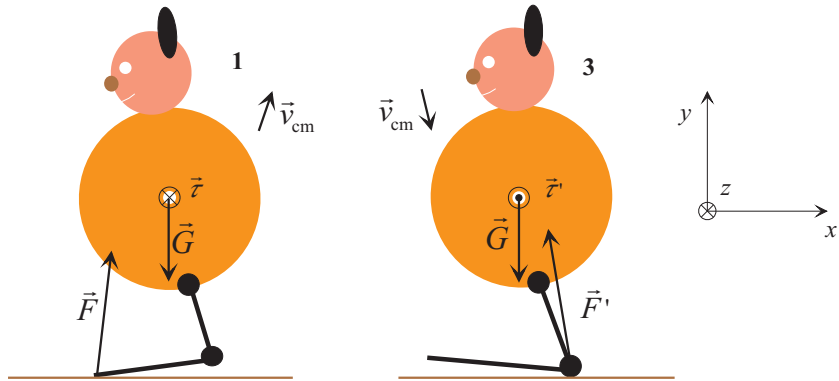}
\end{center}
\caption[]{\label{fig:cang3} \small Constant forces acting on the kangaroo during  the initial (1) and the final (3) phases.  Note that the torques produced by the contact forces are opposite; in phase (1) the torque leads to the rotation of the toy, whereas the torque at phase (3) produces an angular deceleration and eventually the rotation ceases.  }
\vspace*{0.3cm}
\end{figure}

The forces exerted by the ground in phases 1 and 3 are surely time dependent \cite{haugland13}. As a simplifying assumption, we replace these variable forces by constant forces that produce, in each phase, exactly the same
impulse as the real force~\cite{guemez13c}. In phase 1, for instance, the force $\vec F (t)$ exerted by the ground is replaced by the {\em constant} force $\vec F$ such that
$\int_{\Delta t_0} \vec F (t) \, \d t= \vec F \Delta t_0$, where $\Delta t_0$ is the duration of that phase, and similarly for phase 3.
At the end of the initial phase, the center of mass velocity is $\vec v_0$ and, at the beginning of phase 3, the center of mass velocity is $\vec v{\,}'_0$.
Regarding the rotation,  the torque with respect to the center of mass, in phase~1, is also a time dependent function. The torque acts during the time interval $\Delta t_0$, producing a certain variation of the angular momentum of the system (note that now
the moment of inertia of the toy, $I$, also slightly varies because of the legs' articulations; however, this variation is rather small --- the legs are very light in comparison with the rest of the body ---  and we can assertively adopt a constant $I$). Here, our approximation consists in assuming a constant torque such that
$\int_{\Delta t_0} \vec \tau (t) \, \d t= \vec \tau \Delta t_0$, where $\vec \tau$ is a constant vector. The torque provides a clockwise angular acceleration, being $\omega_0$ the angular velocity at the end of phase 1.
In   figure \ref{fig:cang3} we represent the {\em constant} forces and  {\em torques} in phases 1 and 3. The constant torque $\vec \tau{\,}'$ in phase 3 leads to an counterclockwise angular acceleration that reduces to zero the initial  angular velocity ($\omega_0$). Of course, in all phases, the weight, $\vec G$, is always acting, but its torque always vanishes with respect to the center of mass.

The fact that we are considering constant forces and torques considerably simplifies the integrals in the general expressions presented in section 2. These forces and torques are time dependent and they also depend on the position of the centre of mass (the forces) and of the rotation angle, $\phi$ (the torques). The real forces and torques produce certain impulses and angular impulses (right hand sides of equation (\ref{eq1}) and of the first equation (\ref{eq3})). The simplifying assumption consists in replacing the real forces and torques by constant vectors that produce exactly the same impulses. In this way we are simplifying, but not oversimplifying, the problem, making it handy, even practicable in the classroom context.
The constant forces and torques can be regarded as average ones, producing the same momentum (linear and angular) variations as the real ones.
Of course we could use a probably more realistic force such as $F(t)=F_0\sin(\xi t)$, where $\xi$ is a parameter. This would complicate the approach because (1) would no longer be a trivial integration. With a sophisticated
force sensor (the toy is very light) it would be possible, in principle, to figure out $\vec F(t)$ and $\vec F'(t)$ and adjust them with analytic functions. With known time dependent analytic functions the integral in (\ref{eq1}) would be straightforward but the integral in (\ref{eq2}) still would require the knowledge of  $\vec F=\vec F(x_{\rm cm})$ (the same for $\vec F'$) and the integral in (\ref{eq3}) would require the knowledge of $\vec \tau=\vec \tau(\phi)$ (the same for $\vec \tau '$). A more quantitative  analysis of the motion is out of the scope of the present study and this is why we are using constant forces (and torques) to keep the problem within manageable limits.

\begin{figure}[htb]
\begin{center}
\hspace*{0.5cm}
\includegraphics[width=9cm]{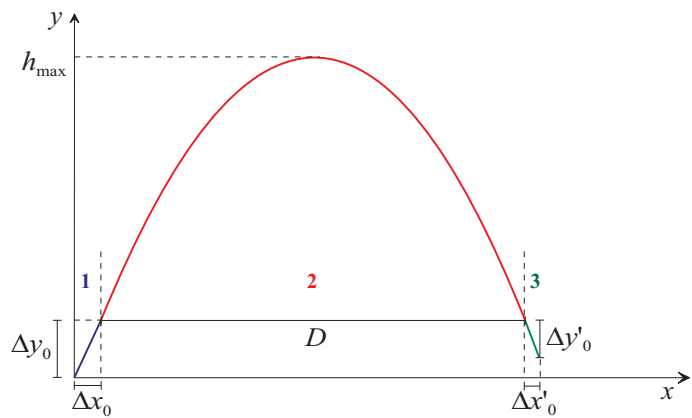}
\end{center}
\caption[]{\label{fig:cang4} \small Kangaroo's center of mass trajectory.  In the first phase it is exactly a straight line, whereas in the second phase it is a parabola. In the third phase it is almost indistinct from a straight line.} 
\end{figure}

Phase 2 consists of a  parabolic motion of the center of mass (neglecting air resistance, of course) combined with an uniform rotation.
In figure \ref{fig:cang4} we show the trajectory of the center of mass of the toy. Assuming constant forces, the trajectory in phase 1 is exactly a straight line.
In fact, for a constant force along $x$ and $y$, and for a body that starts from rest, the accelerations $a_x$ and $a_y$ are constants. Therefore, $x={1\over 2} a_x t^2$ and $y={1\over 2} a_y t^2$, hence, by eliminating $t$ in both equations, this leads to  $y=cx$ ($c$ is a constant), whose graph  is a straight line.
In phase 3
the function $y=y(x)$ is more complicated because there are initial velocities along $x$ and $y$ and $y(x)$ is, in general, not a linear function. However, since the center of mass only moves very little  and for a very short time, the trajectory can be approximated by a straight line (the difference between the straight line --- phase 3 in figure \ref{fig:cang4} --- and the actual trajectory is tiny, even indistinguishable within the precision of the drawing).

In summary, regarding the center of mass motion, it is uniformly accelerated along both $x$ and $y$ in phase 1, it is a projectile motion in phase 2 and, finally, it is
a uniformly retarded motion in phase~3 along $x$ as well as along $y$. Regarding the rotational motion, the angular acceleration is constant in phase 1, in phase 2 the angular velocity is constant, and in phase 3  the angular acceleration is again constant, producing an uniformly retarded angular motion.

\begin{figure}[htb]
\begin{center}
\hspace*{0.5cm}
\includegraphics[width=7.5cm]{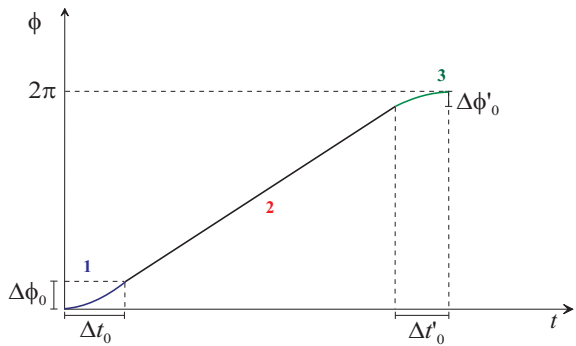}
\end{center}
\caption[]{\label{fig:cang5} \small  Angular displacement around the axis that passes through the kangaroo's center of mass versus time. The relative durations of the initial and final phases are
exaggerated. }
\vspace*{0.2cm}
\end{figure}

Concerning the angular displacement, the toy performs a complete turn around its centre of mass. Most of the time, the toy is in the air, hence the 360$^{\rm o}$ turn is almost executed in phase 2. In phase 1, the supposedly constant torque  produces a quadratic dependence  with time of the angular displacement: $\phi(t) = {1\over 2} \alpha t^2$, where $\alpha$ is the angular acceleration;  the function $\phi(t)$ varies linearly with time during phase 2:
$\phi(t) = \Delta \phi_0 + \omega_0 t$, where $\omega_0$ is the angular velocity at the end of phase 1 and during phase 2; and it again varies quadratically in phase 3:  $\phi(t) = 2\pi-\Delta \phi'_0 + \omega_0 t - {1\over 2} \alpha' t^2$, where $\alpha'$ is the magnitude of the angular acceleration in the third phase. The time, $t$, in the previous expressions starts at the
beginning of each phase. In figure \ref{fig:cang5}  the function $\phi=\phi(t)$ is shown, assuming that the turn corresponds to exactly $2     \pi$ (in practice, this is only an approximate value) and exaggerating, to make the figure more clear, the durations of phases 1 and 3.

\begin{figure}[htb]
\begin{center}
\hspace*{0.5cm}
\includegraphics[width=16cm]{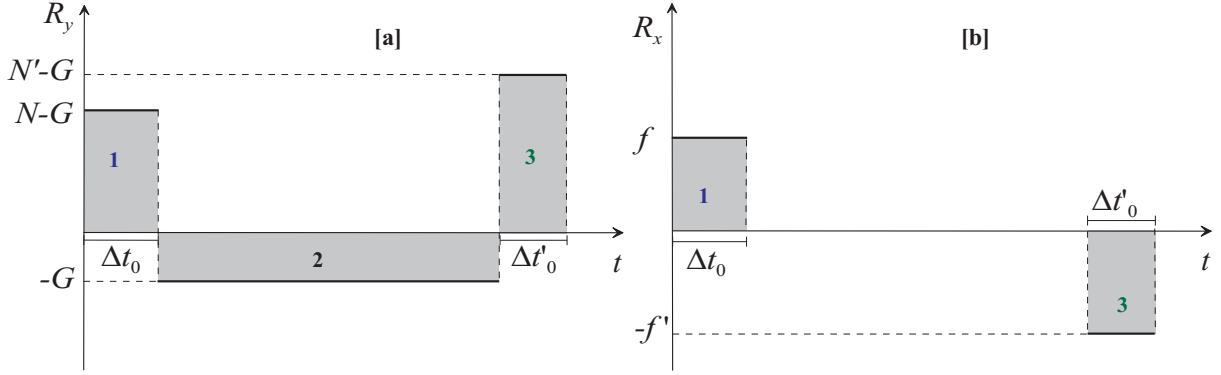}
\end{center}
\caption[]{\label{fig:cang6} \small  Resultant force acting on the toy: vertical component [a], and horizontal component, [b]. The represented areas are the impulses. The impulses, in each graph, add up to zero. The relative durations of the initial and final phases are  exaggerated. There is a picture similar to [b] for the angular impulse (positive in phase 1, negative in phase 3 and null during phase 2). }
\vspace*{0.2cm}
\end{figure}

In addition to our previous assumptions, we consider that the magnitude of the velocities $v_0$ and $v'_0$ is the same, i.e. the kangaroo touches the ground with its center of mass
exactly at the level it occupies when the parabolic motion starts (such an assumption is convenient to simplify the analysis but it could be relaxed).
For the sake of a general discussion we let the displacement $\Delta x_0$ be different from $\Delta x'_0$ (the same for the initial and final $y$ displacements).
 In such a case, the magnitudes of the forces $\vec F$ and $\vec F'$ are different and different are the time intervals during which they act. Indeed, we can stay more general, since such generality has no drastic consequence on the formalism or on its clarity.
So, we assume that the contact constant forces, $\vec F$ and $\vec F'$, do not necessarily have the same magnitude.  In figure \ref{fig:cang6} we represent the vertical component of the resultant force, $R_y$, acting on the toy (part [a]) and the horizontal component of the same resultant force, $R_x$, (part [b]). Again the time intervals in the initial and final phases are exaggerated. The important point to notice is that, for each graph, the algebraic sum of the represented areas, i.e. of the impulses along $y$ and $x$ should add up to zero: the toy stars from rest and it comes to rest at the end of the jump, therefore the total variation of the linear momentum is zero in both the $x$ and $y$ directions.
The same is true for the angular impulses: those in phases (1) and (3) cancel out and, in phase 2, the angular impulse is zero.

Each of the contact forces has a vertical component (normal reaction) and a horizontal component (static friction force).
We may decompose the contact forces according to  $\vec F= f\, \vec{\rm e}_x + N \, \vec{\rm e}_y$ and $\vec F'= -f'\, \vec{\rm e}_x + N' \, \vec{\rm e}_y$. To fix the notation we summarize in table \ref{tab1} the displacements, velocity variations, accelerations and other magnitudes for each phase of the motion.

\begin{table}[bht]
\vskip0.5cm
\begin{center}
\begin{tabular}{|l|| c| c| c| }
\hline
& & & \\
& Phase 1 & Phase 2 & Phase 3 \\
& & & \\
\hline \hline
Duration & $\Delta t_0$ & $\Delta t$ & $\Delta t'_0$ \\
\hline
Force & $(f, N - G)$   &  $(0, -G)$   & $(-f', N'-G) $  \\
\hline
Acceleration & $(a_x,a_y) $ & $ (0, -g)$ &  $(-a'_x,a'_y) $\\
Velocity variation & $(v_{0x},v_{0y})$ &  $(0,-2v_{0y})$ & $(-v_{0x},v_{0y})$  \\
Displacement &  $(\Delta x_{0}, \Delta y_{0})$ & $(D, 0)$ & $(\Delta x'_{0}, - \Delta y'_{0})$ \\
\hline
Torque &   $\tau$ &   $0$  &  $ -\tau'$ \\
\hline
Angular acceleration & $\alpha$ & 0 & $-\alpha '$ \\
Angular velocity variation & $\omega _0$ & 0 & $-\omega _0$ \\
Angular displacement & $\Delta \phi_0 \sim 0^{\rm o}$ & $\sim 360^{\rm o}$ & $\Delta \phi'_0\sim 0^{\rm o}$ \\ \hline

\end{tabular}

\vspace*{0.5cm}
\caption[]{\label{tab1} \small Magnitudes  at each phase of the kangaroo's motion. All quantities are positive, so the negative entries are explicitly
written with a minus sign. The torque is along the $z$ axis, perpendicular to the $xy$ plane. }
\end{center}
\end{table}

Phases (1) and (3) are the most interesting ones, since phase (2) corresponds to the well known projectile motion.
Equation (\ref{eq1}) and the first equation (\ref{eq3}), on the one hand, and equation (\ref{eq2}) and the second equation in (\ref{eq3}), on the other hand, lead to
\beq
{\rm Phase \ 1:} \ \ \ \ \ \left\{\begin{array}{rl} m v_{0x} & = f\, \Delta t_0 \\  m v_{0y} & = (N -G) \, \Delta t_0 \\ I \omega_0 & = \tau \, \Delta t_0
 \end{array} \right.   \ \ \ {\rm and } \ \ \ \ \left\{ \begin{array}{rl}
 {1 \over 2} m v_0^2 &=  f \Delta x_0 + (N-G)\Delta y_0       \\  & \\ {1 \over 2} I \omega_0^2 &=  \tau \Delta \phi_0\, ,
  \end{array}
  \right.
  \label{phase1a}
\eeq
where $m$ is the mass of the toy and $I$ its moment of inertia (that may slightly vary with time but we assume it is constant).
Moreover, equation(\ref{totale}), leads to
\beq
\hspace*{-4.4cm}  {\rm Phase \ 1:} \hspace*{3cm}    {1 \over 2} m v_0^2 +  {1 \over 2} I \omega_0^2  + \Delta U_{\kappa} =  - G \Delta y_0\, ,
 \label{phase1b}
\eeq
where $\Delta U_{\kappa}$ is the part of the internal energy variation due to the spiral spring (this is the potential elastic energy delivered by the spring). The other part of the variation of the internal energy of the system is
the kinetic rotational energy in (\ref{phase1b}).

The corresponding equations for phase 3 are
\beq
{\rm Phase \ 3:} \ \ \ \left\{\begin{array}{rl} m v_{0x} & = f'\, \Delta t'_0 \\  m v_{0y} & = (N' -G) \, \Delta t'_0 \\ I \omega_0 & = \tau' \, \Delta t'_0
 \end{array} \right.   \ \ \ {\rm and } \ \ \ \ \left\{ \begin{array}{rl}
 {1 \over 2} m v_0^2 &=  f' \Delta x'_0 + (N'-G)\Delta y'_0       \\  & \\ {1 \over 2} m \omega_0^2 &=  \tau' \Delta \phi'_0
  \end{array}
  \right.
 \label{phase3a}
 \eeq
and
\beq
\hspace*{-4.8cm} {\rm Phase \ 3:} \hspace*{3cm}   -{1 \over 2} m v_0^2 -  {1 \over 2} I \omega_0^2 =   G \Delta y'_0 + Q\, ,
 \label{phase3b}
 \eeq
where $Q$ is the heat transfer to the surroundings during the process (in this phase $\Delta U_\kappa =0$). We stress that neither $\vec F$ nor $\vec F'$  perform any work because (ideally) their application
points do not move.

As already mentioned, phase 2 corresponds to a projectile motion combined with a uniform rotation. In this part only a conservative force is acting, hence the sum of the translational kinetic energy, of the rotational kinetic energy and of the potential gravitational energy remains constant (the rotational kinetic energy, ${1\over 2} I \omega_0^2$, is,  itself, constant). For this phase, the center of mass equation~(\ref{eq2}) and the first law of thermodynamics (\ref{totale}) provide exactly the same information, namely
\beq
\hspace*{-5.cm} {\rm Phase \ 2:} \hspace*{3cm}   {1 \over 2} m (v^2 - v^2_0) = -  G (y-\Delta y_0)
\eeq
(note that now, in equation (\ref{totale}), $\Delta U =0$ and $Q=0$, the process is a purely mechanical one, because we have neglected the air friction).

Regarding kinematical aspects, the maximum height reached by the center of mass, the horizontal distance traveled by the center of mass and the time of flight are given by
\beq
\hspace*{-2.8cm} {\rm Phase \ 2:} \hspace*{1.5cm} h_{\rm max} = \Delta y_0+{v_{0y}^2 \over 2 g}, \ \ \ \ \ D= {2 v_{0x} \, v_{0y} \over g}, \ \ \ \ \Delta t = {2 v_{oy} \over g}\, .
\label{rts}
\eeq
For this phase, we may write $\Delta \phi = \omega_0 \, \Delta t$. If we take $\Delta \phi \sim 2 \pi$, for the time of flight given in (\ref{rts}), one finds a relation between the vertical component of the velocity at the end of phase 1 and the angular velocity at that very same moment: $\omega_0={\pi g \over v_{0y}}$.

\section{Energetic issues}

In this section we explicitly show that there are asymmetric energetic issues related to the motion in phases 1 and 3, though the mechanical description
of those phases is symmetric in the sense that in the first and in the third phases the impulses of the resultant forces are equal in magnitude with opposite directions.

Let us go back to phase 1. The required  energy for the kangaroo to perform the back somersault is obviously provided by the spiral spring and ultimately by the person that winds it.
When the spring is released, we assume  the elastic energy decrease to be given by $\Delta U_{\kappa}= {1\over 2} \kappa (\theta^2 - \theta^2_0)$,
with ${1\over 2}\kappa \theta_0^2$ the initial stored energy and $\kappa$ the elastic constant characterizing the spring (the angle $\theta$, with initial value $\theta_0$, is a decreasing function of time). As this energy decreases, the articulated toy starts the jump.

Combining equations (\ref{phase1a}) and (\ref{phase1b}), we may express the internal energy variation by
\beqa
\Delta U_{\kappa} & = & - \left(  {1 \over 2} m v_0^2 +  {1 \over 2} I \omega_0^2  + G\Delta y_0  \right) \\ \nonumber
         & = & - \left[   f \Delta x_0 + (N-G)\Delta y_0  + \tau \Delta \phi_0 + G\Delta  y_0 \right] <0\, .
\eeqa
The first line explicitly shows that the internal (elastic) energy is converted into another form of mechanical energy: translational and rotational kinetic energies plus potential gravitational energy.
This phase is reversible --- it evolves with no variation of the entropy of the universe \cite{leff12-iv}.
In the second line, the kinetic energy terms are expressed by the pseudo works related to the contact force and to its torque.

Regarding phase 3,  equations (\ref{phase3a}) and (\ref{phase3b}) now lead to
\beqa
Q & = & - \left(   {1 \over 2} m v_0^2 +  {1 \over 2} I \omega_0^2  + G\Delta y'_0  \right) \\ \nonumber
         & = & - \left[   f' \Delta x'_0 + (N'-G)\Delta y'_0  + \tau' \Delta \phi'_0 + G\Delta  y'_0 \right] <0\, .
\eeqa
This is the heat released in phase 3 of the motion. This energy is simply lost in the sense that it flows from the system to the surroundings, where it is dissipated,
and the entropy of the universe increases~\cite{leff12-v}.
The relation between the original internal energy variation and this heat can be found combining the previous equations, or simply by
applying directly equation (\ref{totale})   to the whole process: on the left hand side of that equation the variation of the center of mass kinetic energy is zero and  the variation of the internal energy is solely
$\Delta U_{\kappa}$
(to simplify the discussion we are  assuming no temperature variations in the toy);
on the right hand side of equation (\ref{totale}), the work is only due to the weight (again we stress that neither the normal forces, nor the static friction forces perform any work);
finally, there is the heat flow  to the surroundings. Altogether, equation (\ref{totale}) yields for the overall process
\beq
Q=\Delta U_{\kappa} + G (\Delta y_0-\Delta y'_0)\, .
\label{caloor}
\eeq
For $\Delta y_0=\Delta y'_0$
the energy initially stored in the spring totally dissipates in the surroundings, which behaves as a heat reservoir at temperature $T$.  If the final position of the center of mass lies at an higher level than the initial one, part of the initial stored energy is used
to raise the center of mass of the kangaroo, as expressed in equation (\ref{caloor}). In that case, the magnitude of the heat is smaller than the magnitude of the elastic potential energy initially stored in the toy.
The overall process is clearly irreversible and, for  $\Delta y_0=\Delta y'_0$, the entropy variation of the universe is $\Delta S_{\rm U}= -{Q \over T}= -{\Delta U_\kappa \over T}>0$ (the heat should now be considered from the point of view
of the reservoir, and this is the reason for the minus sign).

When $\Delta x_0=\Delta x'_0$ and  $\Delta y_0=\Delta y'_0$, even though the mechanical analysis becomes quite symmetric for the initial and final phases, there is a clear thermodynamical asymmetry: in phase 1, there is mechanical energy production at the expend of the internal energy of the system; in phase 3, the mechanical energy is ``lost" in the sense that it dissipates as heat in the surroundings and it cannot be recovered in an useful way.
We may say that, at the end of phase 1, the system still keeps the energy in an ``organized form" as mechanical energy (hence, no entropy increase), whereas in the final phase there is no  mechanism to recover the decrease of mechanical energy: in particular that energy cannot be stored back in the spiral spring. If that could be possible, the kangaroo would perform back somersaults continuously... We all know that this is not the case: after one jump one has to
rewind the spiral spring for the next jump.

In a recent publication we discuss the same type of asymmetries in the context of human movements, like jumping and walking \cite{guemez14b}. As in the example discussed in this paper, in one part of the
process there is a direct conversion of internal energy into mechanical energy but, in the other part, the mechanical energy dissipates as heat in the surroundings.

  \section{Conclusions}

The motion of the kangaroo studied in this paper can be used in the classroom to explicitly show the correspondence between translations and rotations. The example also serves to demonstrate that, for certain mechanical systems performing movements in which one part is symmetric with respect to the other, the thermodynamical behavior of both parts may be different. More precisely, in one part there may be production of mechanical energy, due to an internal energy source, a process that does not increase the entropy, whereas in the other part of the motion there occurs dissipation of mechanical energy, which is transferred to the surroundings as heat, and there is an unescapable entropy increase.

We should emphasize that the first phase of the problem studied in this paper is somehow similar to the springboard diver discussed in ref.~\cite{walker11} --- the main difference is that the internal (elastic) energy variation
should then be replaced by the variation of the Gibbs free energy  in the person's muscles \cite{guemez14b,atkins10}. There are of course many other examples but the point we would like to emphasize here is the fact that energetic aspects are usually a bit underrated in  textbooks
\cite[p.~70]{walker11}.

The discussion presented in this paper comes in the sequence of a series of papers~\cite{guemez13,guemez14,guemez13b,guemez13c,guemez14b} devoted to the link between mechanics and thermodynamics, a tendency that, fortunately, is already
present in modern textbooks \cite{chabay11x}. Our aim is to contribute to fill the  gap one encounters in most treatments of classical mechanics, and we do hope that our discussion, motivated by a ``kangaroo",
is relevant for physics education.


\end{document}